\documentclass[12pt]{article}
\usepackage{citesort,fullpage,epsfig,psfrag,graphics,amsbsy,amssymb,amsmath
}
\usepackage{caption}
\newcommand{\beq}{\begin{equation}}
\newcommand{\eeq}{\end{equation}}
\newcommand{\bea}{\begin{eqnarray}}
\newcommand{\eea}{\end{eqnarray}}
\newcommand{\bfs}{\boldsymbol}

\newcommand{\be}{\begin{equation}}
\newcommand{\ee}{\end{equation}}
\newcommand{\bq}{\begin{eqnarray}}
\newcommand{\eq}{\end{eqnarray}}

\def\math{\mathsurround=0pt }
\def\leftrightarrowfill{$\math \mathord\leftarrow \mkern-6mu 
 \cleaders\hbox{$\mkern-2mu \mathord- \mkern-2mu$}\hfill
 \mkern-6mu \mathord\rightarrow$}
\def\overleftrightarrow#1{\vbox{\ialign{##\crcr
     \leftrightarrowfill\crcr\noalign{\kern-1pt\nointerlineskip}
     $\hfil\displaystyle{#1}\hfil$\crcr}}}

\let\l=\lambda

 \def\bd{\begin{document}} \def\ed{\end{document}}
\def\ds{\documentstyle} \let\fr=\frac \let\bl=\bigl \let\br=\bigr
\let\Br=\Bigr \let\Bl=\Bigl
\let\bm=\bibitem
\let\na=\nabla
\let\pa=\partial \let\ov=\overline
\def\ft#1#2{{\textstyle{{\scriptstyle #1}\over {\scriptstyle #2}}}}
\def\fft#1#2{{#1 \over #2}}
\def\vp{\varphi}
\def\sst#1{{\scriptscriptstyle #1}}
\def\oneone{\rlap 1\mkern4mu{\rm l}}
\def\td{\tilde}
\def\wtd{\widetilde}
\def\dalemb#1#2{{\vbox{\hrule height .#2pt
        \hbox{\vrule width.#2pt height#1pt \kern#1pt
                \vrule width.#2pt}
        \hrule height.#2pt}}}
\def\square{\mathord{\dalemb{6.8}{7}\hbox{\hskip1pt}}}
\def\wtd{\widetilde}
\def\R{\rlap{\rm I}\mkern3mu{\rm R}}
\def\im{{\rm i}}
\def\tilg{\tilde{g}}
\def\tilF{\tilde{F}}
\def\tilA{\tilde{A}}
\def\varf{\varphi}
\def\tilf{\tilde{\phi}}
\def\tilh{\tilde{h}}
\def\rme{{\rm e}}
\def\ep{\epsilon}
\def\0{{(0)}}
\def\9{{(9)}}
\def\8{{(8)}}
\def\7{{(7)}}
\def\6{{(6)}}
\def\5{{(5)}}
\def\4{{(4)}}
\def\3{{(3)}}
\def\2{{(2)}}
\def\1{{(1)}}
\newcommand{\trace}{{\rm Tr}}
\newcommand{\ub}{\overline{U}}
\newcommand{\vb}{\overline{V}}
\newcommand{\uh}{\widehat{U}}
\newcommand{\vh}{\widehat{V}}
\newcommand{\ubh}{\overline{\widehat{U}}}
\newcommand{\vbh}{\overline{\widehat{V}}}
\newcommand{\lb}{\bar{\l}}
\newcommand{\Fb}{\overline{F}}
\newcommand{\Fh}{\widehat{F}}
\newcommand{\Fbh}{\overline{\widehat{F}}}
\newcommand{\Ab}{\overline{A}}
\newcommand{\Ah}{\widehat{A}}
\newcommand{\Abh}{\overline{\widehat{A}}}
\newcommand{\Gb}{\overline{G}}
\newcommand{\Gh}{\widehat{G}}
\newcommand{\Gbh}{\overline{\widehat{G}}}
\newcommand{\Pb}{\overline{P}}
\newcommand{\Ph}{\widehat{P}}
\newcommand{\Pbh}{\overline{\widehat{P}}}
\newcommand{\Qb}{\overline{Q}}
\newcommand{\Qh}{\widehat{Q}}
\newcommand{\Qbh}{\overline{\widehat{Q}}}
\newcommand{\Bb}{\overline{B}}
\newcommand{\Bh}{\widehat{B}}
\newcommand{\Bbh}{\overline{\widehat{B}}}
\newcommand{\fhns}{\hat{F}^{\rm (NS)}}
\newcommand{\fhrr}{\hat{F}^{\rm (RR)}}
\newcommand{\ahns}{\hat{A}^{\rm (NS)}}
\newcommand{\ahrr}{\hat{A}^{\rm (RR)}}
\newcommand{\hhrr}{\hat{H}^{\rm (RR)}}
\newcommand{\hchi}{\hat{\chi}}
\newcommand{\hphi}{\hat{\phi}}
\newcommand{\htau}{\hat{\tau}}
\newcommand{\cG}{{\cal G}}
\newcommand{\cGb}{\overline{{\cal G}}}
\newcommand{\cH}{{\cal H}}
\newcommand{\cP}{{\cal P}}
\newcommand{\cPb}{\overline{{\cal P}}}
\newcommand{\cQ}{{\cal Q}}
\newcommand{\cQb}{\overline{{\cal Q}}}
\newcommand{\cM}{{\cal M}}
\newcommand{\cN}{{\cal N}}
\newcommand{\cO}{{\cal O}}
\newcommand{\cD}{{\cal D}}
\newcommand{\cL}{{\cal L}}

\newcommand{\vpp}{\mbox{$\langle{\scriptstyle++}\rangle$}}
\newcommand{\vmp}{\mbox{$\langle{\scriptstyle-+}\rangle$}}
\newcommand{\vppp}{\mbox{$\langle{\scriptstyle+++}\rangle$}}
\newcommand{\vmpp}{\mbox{$\langle{\scriptstyle-++}\rangle$}}
\newcommand{\vpmp}{\mbox{$\langle{\scriptstyle+-+}\rangle$}}
\newcommand{\goesas}[1]{{}_{{\displaystyle\sim}\atop#1}}
\newcommand{\impliesas}[1]{{}_{{\displaystyle{\Longrightarrow}}\atop#1}}
\renewcommand{\thepage}{\arabic{page}}

\begin{document}
\setlength{\captionmargin}{36pt}
\begin{titlepage}
\begin{flushright}
\phantom{UFIFT}
\end{flushright}

\vskip 3cm
\begin{center}
\begin{large}
{\bf Four String Amplitudes for the Generalized Protostring.}
\end{large}
\vskip 2cm
{\large 
Charles B. Thorn
\footnote{E-mail  address: {\tt thorn@phys.ufl.edu}}
}
\vskip0.20cm
{\it Institute for Fundamental Theory,\\
Department of Physics, 
University of Florida
Gainesville FL 32611
}

\vskip24pt
\end{center}
\begin{abstract}
  \noindent
  We specialize the $N$ string scattering amplitudes for
  the generalized protostring to $N=4$. This allows
  for a much more detailed and explicit study of their basic
  physical and mathematical properties, such as singularity structure
  and high energy behavior. Since this class of models does
  not enjoy full Poincar\'e invariance, the high energy behavior
  depends on the Lorentz frame. The relative simplicity of the four
  string amplitudes allows a direct understanding of the complications
  due to this non-covariance.

\end{abstract}

\vfill
\end{titlepage}
\section{Introduction}
Several years ago, motivated by string bit models
\cite{gilest,thornsakh,bergmantsubit}, some novel string theories, including
ones which lacked full Poincar\'e invariance, were proposed
\cite{thornprotobits}. Despite the string bit motivation,
these models can be formulated directly on the continuous
lightcone worldsheet \cite{goddardrt,goddardgrt}. In particular,
Mandelstam's interacting string formalism 
\cite{mandelstamlc,mandelstamdet}
could be used to obtain multi-string
scattering amplitudes \cite{thornprotoamp}. Although
formulas for these
amplitudes were obtained, their physical properties were not studied
extensively. This paper is an addendum to \cite{thornprotoamp} focusing
exclusively on the mathematical and physical
properties of four open string amplitudes in these models.


The protostring model can be defined as the string model in which 
each of the 24 transverse coordinate worldsheet fields 
of the lightcone bosonic string is replaced by a spinor valued  
left-right pair of integer moded
Grassmann worldsheet fields $\theta_L,\theta_R$.
In \cite{thornprotoamp}  we obtained amplitudes for a
generalization of the protostring in which only $s$ bosonic dimensions are 
so replaced,
with the remaining $d=24-s$ left as transverse coordinates. 
In all these models, the
string interaction is a simple overlap without operator insertions
at the join/break point. The condition $s+d=24$ ensures the finite
continuum limit of the string bit overlap. 

With the continuum scattering
amplitude written in the form ${\cal M}\prod_k|p_k^+|^{-1/2}$, this finiteness
condition means that ${\cal M}$ is invariant under the scale transformation
$p_k^+\to\lambda p_k^+$. In other words, invariance under
the subgroup $SO(1,1)$ of the Lorentz group in $d+1$ space dimensions
($SO(d+1,1)$) is maintained. For the protostring ($s=24$ or $d=0$), 
this is the entire Lorentz group, but for $s<24$,  with the exception of the
bosonic string ($s=0$), the Lorentz group is broken
to $SO(1,1)\times SO(d)$ \footnote{The $SO(d)$ factor can be extended to
  the Galilei group in $d$ space dimensions by including the
  Galilei boosts $M^{i+}$, with $i=1,\ldots,d$, of lightcone quantization.
  The broken generators are $M^{i-}$ when $d<24$ or $s>0$.}.

We assume here, as in \cite{thornprotoamp}, that $s$ is even
so the Grassmann worldsheet fields may be replaced by
$s/2$ compactified bosonic worldsheet fields $\phi^r$. Integer moded Grassmann
fields are equivalent to compactified bosonic fields for which the
components of the momenta ${\bfs\pi}_k$ are quantized
in odd multiples of a fixed number $\gamma$
with $\gamma^2=1/8$ for the open string and $\gamma^2=1/2$ for the
closed string. By comparing the overlap in the original spinor formulation
to the bosonized formulation, Sun showed that the
three string vertex violates momentum conservation
by $\pm\gamma$ for each of the $s/2$ components \cite{songge}.
The bosonized overlap
therefore contains operator factors $\prod_r\cos(\gamma\phi_r)$,
which insert, for each component, momentum $\pm\gamma$ into the process.
For $N$ string amplitudes, there are $N-2$ such factors, which are
described by $N-2$ $s/2$-vectors ${\bfs\gamma}_r$ which satisfy
the constraint
\bea
\sum_k{\bfs\pi}_k+\sum_r{\bfs\gamma}_r&=&0.
\label{constraint}
\eea
For even $N$ it is possible that $\sum_r{\bfs\gamma}_r=0$,
allowing nonzero momentum-conserving scattering amplitudes.

There are also the momenta ${\bfs p}_k$ of the $d=24-s$
uncompactified transverse coordinates and their Minkowski extensions
$p^\mu=({\bfs p}_k,p^+_k,p^-_k)$. Finally it is convenient to append
to $p^\mu$ the ${\bfs\pi}$ and denote the resulting $2+d+s/2$ Minkowski
vector by $P^\mu$. In this notation the mass shell condition is,
in units where $\alpha^\prime=1$,
$P\cdot P\equiv {\bfs p}^2+{\bfs\pi}^2-2p^+p^-=1$ for each external string.

The scattering amplitudes of \cite{thornprotoamp}, quoted
for the reader's convenience in the appendix,
involve
$N$ external strings, all with no oscillator excitations
and, in the closed string case, with zero winding number. However,
their compactified momenta ${\bfs\pi}_k$ could
have components any odd multiple of $\pm\gamma$.
In this paper, we specialize further to external strings of minimal mass.
Then each component of each  ${\bfs \pi}_k$
should be $\pm\gamma$. The $d+2$ dimensional mass squared of each string
is then $-p\cdot p=-(P^2-{\bfs\pi}^2)=-1+{\bfs\pi}^2=s/12-1$
for the open string, or $s/3-4$ for the closed string.
There are all together
$N+N-2=2(N-1)$ ${\bfs\gamma}$'s and ${\bfs\pi}$'s, so to satisfy the 
conservation law 
$N-1$ of them should have value $+\gamma$ and the remaining ones
should have value
$-\gamma$. There are $\genfrac{(}{)}{0pt}{1}{2(N-1)}{N-1}$ ways to do this
for each of the $s/2$ compactified bosonic fields. For $N=4$ this means that
there are $20^{s/2}$ ways to assign $\pm\gamma$ to the ${\bfs\pi}$'s
and ${\bfs\gamma}$'s.

In the next Section 2 we discuss the conformal mappings required to evaluate
any 4 string amplitude.
Section 3 studies the amplitudes for several simple choices for
the ${\bfs\pi}_k$'s. Section 4 analyzes high energy scattering in the
2 to 2 case. Concluding comments are in Section 5. An appendix
collects formulas for the $N$-string scattering amplitudes obtained in
\cite{thornprotoamp}.
\section{Conformal maps for four string amplitudes}
First specialize the conformal mapping from the complex $z$-plane
to the lightcone worldsheet $\rho$ to four
external strings \cite{mandelstamlc}:
\bea
\rho&=&\alpha_1\ln z+\alpha_2\ln(z-Z)+\alpha_3\ln(z-1).
\label{ztorho4}\eea
Then the two interaction points are determined by $d\rho/dz=0$,
which implies 
\bea
x_\pm&=&\frac{\alpha_1(Z+1)+\alpha_2+\alpha_3Z\pm
\sqrt{(\alpha_1(Z+1)+\alpha_2+\alpha_3Z)^2+4\alpha_1\alpha_4Z}
}{2(\alpha_1+\alpha_2+\alpha_3)}\\
&=&\frac{\alpha_{12}+\alpha_{13}Z\pm
\sqrt{\alpha_{12}^2+\alpha_{13}^2Z^2
+2(\alpha_1\alpha_4+\alpha_2\alpha_3)Z}}{-2\alpha_4}
\\
&=&1+\frac{\alpha_{14}-\alpha_{13}(1-Z)\pm
  \sqrt{\alpha_{14}^2+\alpha_{13}^2(1-Z)^2+2(\alpha_1\alpha_2+\alpha_3\alpha_4)
    (1-Z)}}{-2\alpha_4}.
\label{exes}
\eea
The alternative forms follow from the identities
\bea
\alpha_4^2|x_+-x_-|^2
&=&\alpha_{12}^2(1-Z)+\alpha_{23}^2Z
-\alpha_{13}^2Z(1-Z)\label{diffx1}\\
&=&\alpha_{12}^2+\alpha_{13}^2Z^2
+2(\alpha_1\alpha_4+\alpha_2\alpha_3)Z\label{diffx2}\\
&=&\alpha_{14}^2+\alpha_{13}^2(1-Z)^2
+2(\alpha_1\alpha_2+\alpha_3\alpha_4)(1-Z),
\label{diffx3}
\eea
where $\alpha_{kl}\equiv \alpha_k+\alpha_l$.
The last two lines (\ref{diffx2}) and (\ref{diffx3}),
give forms convenient for the study of the $Z\sim0$
and $Z\sim1$ limits respectively.

The integrand for the 4 string amplitude requires the factors
\bea
\prod_{r<s}|x_r-x_s|^{2{\bfs\gamma}_r\cdot{\bfs\gamma}_s-s/24}&=&
|x_1-x_2|^{({\bfs\gamma}_1+{\bfs\gamma}_2)^2-s/6}\\
\prod_{k< l<N}|Z_k-Z_l|^{2P_k\cdot P_l-s/24}&=&Z^{-S+({\bfs\pi}_1
  +{\bfs\pi}_2)^2-2}(1-Z)^{-t+({\bfs\pi}_2+{\bfs\pi}_3)^2-2}\\
\prod_{r,k<N}|x_r-Z_k|^{s/24}&=&\left[\frac{\alpha_1\alpha_2\alpha_3}{
  \alpha_4^3}Z^2(1-Z)^2\right]^{s/24}\\
\prod_{r,k<N}|x_r-Z_k|^{2{\bfs\pi}_k{\bfs\gamma}_r}&=&
\prod_{k=1}^3\left|\frac{x_1-Z_k}{x_2-Z_k}\right|^{{\bfs\pi}_k\cdot
  ({\bfs\gamma}_1-{\bfs\gamma}_2)}\prod_{k=1}^3\left|(x_1-Z_k)(x_2-Z_k)
\right|^{{\bfs\pi}_k\cdot
  ({\bfs\gamma}_1+{\bfs\gamma}_2)}\nonumber\\
&=&\left|\frac{x_1}{x_2}\right|^{{\bfs\pi}_1\cdot
  ({\bfs\gamma}_1-{\bfs\gamma}_2)}
\left|\frac{x_1-Z}{x_2-Z}\right|^{{\bfs\pi}_2\cdot
  ({\bfs\gamma}_1-{\bfs\gamma}_2)}
\left|\frac{x_1-1}{x_2-1}\right|^{{\bfs\pi}_3\cdot
  ({\bfs\gamma}_1-{\bfs\gamma}_2)}\nonumber\\
&&\hskip-18pt\left|Z\frac{\alpha_1}{\alpha_4}\right|^{{\bfs\pi}_1\cdot
  ({\bfs\gamma}_1+{\bfs\gamma}_2)}
\left|Z(1-Z)\frac{\alpha_2}{\alpha_4}\right|^{{\bfs\pi}_2\cdot
  ({\bfs\gamma}_1+{\bfs\gamma}_2)}
\left|(1-Z)\frac{\alpha_3}{\alpha_4}\right|^{{\bfs\pi}_3\cdot
  ({\bfs\gamma}_1+{\bfs\gamma}_2)}
\eea
where we have chosen $Z_1=0$, $Z_2=Z$, and $Z_3=1$, and we have introduced
the Mandelstam invariants $S=-(p_1+p_2)^2$ and $t=-(p_2+p_3)^2$.
To construct the open 4-string amplitude one integrates the
product of all these factors over $Z$ from 0 to 1, and sums over
all choices for ${\bfs\gamma}_{1,2}$ consistent with the constraint
(\ref{constraint}).
Note however that for a given component, if $\gamma_1+\gamma_2=\pm2$, then
$\gamma_1-\gamma_2=0$, and vice versa.

\subsection{Crossing Symmetry}
Because there is explicit dependence of the integrand of the amplitude on
the $x_\pm$ when $s\neq 0$, crossing symmetry ($S\leftrightarrow t$
for the open string) is not as transparent as for the bosonic string.
However, it still follows from the simple change of variables $Z\to1-Z$.
This is because of the identities
\bea
x_\pm(1-Z;\alpha_1,\alpha_2,\alpha_3,\alpha_4)
&=&1-x_\mp(Z;\alpha_3,\alpha_2,\alpha_1,\alpha_4)\nonumber\\
1-x_\pm(1-Z;\alpha_1,\alpha_2,\alpha_3,\alpha_4)
&=&x_\mp(Z;\alpha_3,\alpha_2,\alpha_1,\alpha_4)\label{crossing}\\
1-Z-x_\pm(1-Z;\alpha_1,\alpha_2,\alpha_3,\alpha_4)
&=&-(Z-x_\mp(Z;\alpha_3,\alpha_2,\alpha_1,\alpha_4)).\nonumber
\eea
When we construct the amplitudes in the next section,
we shall see that
these identities enable one to show that the amplitudes
are symmetric under
the combined transformation $S\leftrightarrow t$,
${\bfs\pi}_1\leftrightarrow{\bfs\pi}_3$,
and $\alpha_1\leftrightarrow \alpha_3$. Or more simply
under $P_1\leftrightarrow P_3$.

\subsection{Singularities of Amplitudes}
The poles in $-S=(p_1+p_2)^2$ are controlled by $Z\sim0$ and those in
$-t=(p_2+p_3)^2$ by $Z\sim1$. We therefore need the behavior of $x_\pm$
in these regions of integration:
\bea
x_+&\sim&-\frac{\alpha_{12}}{\alpha_4}-\frac{\alpha_2\alpha_3}{\alpha_{12}
  \alpha_4}Z+O(Z^2)\\
x_-&\sim&\frac{\alpha_1}{\alpha_{12}}Z+O(Z^2)  
\label{limitexes0}
\eea
for $Z\sim0$ at fixed $\alpha_k$. and
\bea
x_+&\sim&1-\frac{\alpha_{14}}{\alpha_4}-\frac{\alpha_1\alpha_2}{\alpha_{14}
  \alpha_4}(1-Z)+O((1-Z)^2)\\
x_-&\sim&1+\frac{\alpha_3}{\alpha_{14}}(1-Z)+O((1-Z)^2)
\label{limitexes}\eea
for $Z\sim1$
at fixed $\alpha_k$, and assuming for definiteness, that
$\alpha_{14}=-\alpha_{23}>0$. With the opposite signs the roles of $x_+,x_-$
are switched. As long as $\alpha_{12}$ and $\alpha_{23}$
are both non zero, the lowest lying pole locations are determined
by the factors $x_-$ and $Z-x_-$ when analyzing $Z\sim0$,
and by the factors $1-x_-$ and $Z-x_-$ when $Z\sim1$.
\subsection{High Energy}
The high energy Regge behavior (large $-S$ at fixed $t$) is also controlled by
$Z\sim1$, more precisely by $1-Z\sim1/(-S)$. However, in order to
confirm Mueller's argument \cite{mueller}
for constant cross sections, we should also like to consider
large $-S$ at fixed $\alpha_1/(-S)$ and fixed $\alpha_2,\alpha_3$. This means
that factors like $\alpha_1(1-Z)$
will be of order 1.
Then in leading order, we can approximate the $x_\pm$ by
\bea
x_\pm&\approx&1+\frac{\alpha_{14}-\alpha_{1}(1-Z)\pm
  \sqrt{\alpha_{14}^2+\alpha_{1}^2(1-Z)^2+2(\alpha_1(\alpha_2-\alpha_3)
    (1-Z)}}{-2\alpha_4}.
\label{limitexeshe}
\eea
To analyze the amplitudes in this limit, it is convenient to change
variables $Z=e^{-(u/(-S))}$ and replace $1-Z\approx u/(-S)$. Then
$dZ=-Zdu/(-S)$. Also, to reduce clutter we introduce the ratio
$\eta\equiv \alpha_1/(-S)$. Then the approximate form for $x_\pm$ can be
written
\bea
x_\pm&\approx&1+\frac{\alpha_{14}-\eta u\pm
  \sqrt{\alpha_{14}^2+\eta^2u^2+2\eta u(\alpha_2-\alpha_3)
   }}{-2\alpha_4}.
\label{xhe}\eea
We note that, because the numerator of the second term on the right
is of order 1 in the limit, the second term
is of order $1/\alpha_1$ and can be neglected except in factors such as
$x_\pm-1$ or $x_\pm-Z$ which are of order $1/\alpha_1$ in the limit.
Both amplitudes depend on the $x_\pm$ through the ratios
\bea
\frac{x_-}{x_+}&\approx& 1
\label{ratiohe1}\\
  \frac{x_--1}{x_+-1}&\approx&\frac{\alpha_{14}-\eta u-
  \sqrt{\alpha_{14}^2+\eta^2u^2+2\eta u(\alpha_2-\alpha_3)
   }}{\alpha_{14}-\eta u+
  \sqrt{\alpha_{14}^2+\eta^2u^2+2\eta u(\alpha_2-\alpha_3)
  }}
\label{ratiohe2}\\
\frac{x_--Z}{x_+-Z}&\approx&\frac{\alpha_{14}+\eta u-
  \sqrt{\alpha_{14}^2+\eta^2u^2+2\eta u(\alpha_2-\alpha_3)
   }}{\alpha_{14}+\eta u+
  \sqrt{\alpha_{14}^2+\eta^2u^2+2\eta u(\alpha_2-\alpha_3)
  }}
\label{ratiohe3}
\eea
all of which are of order 1 in the limit,
provided $\eta$ is nonzero and finite.

\section{Amplitudes for four strings with minimal mass}
Each of the $s/2$ components of the four compactified momenta ${\bfs\pi}_k$
is constrained by
\bea
\pi_1^a+\pi_2^a+\pi_3^a+\pi_4^a+\gamma_1^a+\gamma_2^a&=&0.
\label{pigammaconstraint}
\eea
Since we are specializing to minimal mass, these constraints must be satisfied
with each component of ${\bfs\pi}_k$ and ${\bfs\gamma}_k$ either $+\gamma$
or $-\gamma$: three of the six one sign and the other three the opposite
sign. We can categorize the choices by the values of
$\gamma_1^a+\gamma_2^a$: $+2\gamma$, $0$, or $-2\gamma$. The multiplicities
of each choice are $1$, $2$, and $1$ respectively. But these choices can be
made independently for each component. For each external
string, we can write $\pi^a=h^a\gamma$ and refer  to $h^a$ as the
$a$th component of the helicity of that string. Thus we can refer to
the three categories as $\Delta h=+2$, $\Delta h=0$, and $\Delta h=-2$.
In the following we shall concentrate on the cases where all components
are in the same category. But of course there are many ``mixed''
possibilities, which will not be discussed explicitly.
\subsection{$\Delta h=\pm2$}
Putting $N=4$ in (\ref{equalgammao}) we find the four open string amplitude
in the special case $\gamma_r=\gamma=1/\sqrt{8}$, $\pi_k=-\gamma=-1/\sqrt{8}$ 
for $k<4$:
\bea
{\cal M}^{\Delta h=2}&=&
\int_0^1 dZ
\frac{|\alpha_4|^{s/4}|x_2-x_1|^{s/12}}{|\alpha_1\alpha_2\alpha_3|^{s/12}}
Z^{(p_1+p_2)^2-2+s/12}(1-Z)^{(p_2+p_3)^2-2+s/12}\nonumber\\
&=&
\frac{|\alpha_4|^{s/4}}{|\alpha_1\alpha_2\alpha_3|^{s/12}}
\int_0^1 dZZ^{(p_1+p_2)^2-2+s/12}(1-Z)^{(p_2+p_3)^2-2+s/12}\nonumber\\
&&\left|\frac{\alpha_{12}^2+\alpha_{13}^2Z^2
    +2(\alpha_1\alpha_4+\alpha_2\alpha_3)Z}{\alpha_4^2}\right|^{s/24}
\label{delta2amp}
\eea
The expression inside the absolute value signs on the last line stays positive
throughout the integration region $0<Z<1$, so that it is safe to drop them.
It is noteworthy that when $s=24$ (the protostring case) this factor is a
second order polynomial in $Z$,
so the amplitude reduces to a linear combination 
of three Euler beta functions. In addition to this simplification, the
kinematics of scattering is limited to forward and backward scattering
since there is only 1 space dimension when $d=0$. The consequences of these
simplifications were already discussed in \cite{thornprotoamp} and will not be
elaborated further here.

It is not hard to check that the pole singularities in
$s$ and $t$ are where they should be
as long as $\alpha_{12}$ and $\alpha_{23}$ are non zero. This is
reasonable since
excluding these values of the $\alpha$'s guarantees that the 
dynamical singularities are all due to the long time propagation 
of protostring mass eigenstates.
On the other hand, if $\alpha_{23}=0$, $\alpha_4^2(x_2-x_1)^2
\sim4\alpha_1\alpha_2(1-Z)$ as $Z\to1$ so the poles in $t$ are
shifted by an amount
$s/24$. When $\alpha_{23}=0$ these singularities are due to the collision
of the interaction points on the worldsheet and not the long time
propagation of a particle state. This nonuniformity of singularity
structure is absent for the bosonic and superstring because the
amplitude integrands do not depend explicitly on the $x_r$. 
\subsection{$\Delta h^a=0$}
In this case the ${\bfs\pi}$ are conserved, so that
the ${\bfs\gamma}_r$ must sum to zero.
In the 4-string case this requires ${\bfs\gamma}_2=-{\bfs\gamma}_1$. Then the
various factors in the integrand, for each choice for ${\bfs\gamma}_1$, are
\bea
\prod_{r<s}|x_r-x_s|^{2{\bfs\gamma}_r\cdot{\bfs\gamma}_s-s/24}&=&
|x_1-x_2|^{-2{\bfs\gamma}_1^2-s/24}=|x_1-x_2|^{-s/6}\\
\prod_{k< l<N}|Z_k-Z_l|^{2P_k\cdot P_l-s/24}&=&Z^{-S+({\bfs\pi}_1
  +{\bfs\pi}_2)^2-2}(1-Z)^{-t+({\bfs\pi}_2+{\bfs\pi}_3)^2-2}\\
\prod_{r,k<N}|x_r-Z_k|^{s/24}&=&\left[\frac{\alpha_1\alpha_2\alpha_3}{
  \alpha_4^3}Z^2(1-Z)^2\right]^{s/24}\\
\prod_{r,k<N}|x_r-Z_k|^{2{\bfs\pi}_k{\bfs\gamma}}&=&
\prod_{k=1}^3\left|\frac{x_1-Z_k}{x_2-Z_k}\right|^{2{\bfs\pi}_k\cdot
  {\bfs\gamma}_1}\nonumber\\
&=&\left|\frac{x_1}{x_2}\right|^{2{\bfs\pi}_1\cdot
  {\bfs\gamma}_1}
\left|\frac{x_1-Z}{x_2-Z}\right|^{2{\bfs\pi}_2\cdot
  {\bfs\gamma}_1}
\left|\frac{x_1-1}{x_2-1}\right|^{2{\bfs\pi}_3\cdot
  {\bfs\gamma}_1}
\eea
where we have again chosen $Z_1=0$, $Z_2=Z$, and $Z_3=1$.
The complete (tree) amplitude is constructed by taking
the product of all these
factors, summing over all $2^{s/2}$ possibilities for ${\bfs\gamma}_1$ and
integrating $Z$ from 0 to 1. The only factors that
depend on ${\bfs\gamma}_1$ are those on the last line. Since the
components of ${\bfs\gamma}_1$ are $\pm\gamma$,
this sum yields the product of factors
\bea
\prod_{i=1}^{s/2}\left(\left|\frac{x_1}{x_2}\right|^{2{\pi}^i_1
  {\gamma}}
\left|\frac{x_1-Z}{x_2-Z}\right|^{2{\pi}^i_2{\gamma}}
\left|\frac{x_1-1}{x_2-1}\right|^{2{\pi}^i_3{\gamma}}
+(x_1\leftrightarrow x_2)\right)
\eea
where $i$ labels the component of ${\bfs\pi}_k$.

We analyze two simple examples of
fully elastic scattering amplitudes. 
First, we choose ${\bfs\pi}_4=-{\bfs\pi}_1$ {\it and}
${\bfs\pi}_3=-{\bfs\pi}_2$ so that the outgoing strings are
in the same internal states as the incoming ones.
Necessarily then we must have ${\bfs\gamma}_2=-{\bfs\gamma}_1$,
which we are assuming in this subsection anyway.
Let ${\bfs\gamma}$ be the $s/2$ vector with each component
equal to $\gamma=1/\sqrt{8}$ for the open string. Then our first
simple example of elastic scattering is
\bea
{\bfs\pi}_1={\bfs\pi}_2={\bfs\gamma},\qquad {\bfs\pi}_3={\bfs\pi}_4
=-{\bfs\gamma},\qquad {\bfs\gamma}_1=-{\bfs\gamma}_2
\eea
As already mentioned, there are $2^{s/2}$
choices for ${\bfs\gamma_1}$ since each component
can be $\pm\gamma$ independently. The total amplitude
should include the sum over all
such choices.
After summing over both signs for each component
of ${\bfs\gamma}_1$, the total amplitude becomes
\bea
{\cal M}^{\Delta h=0}_1&=&\left[\frac{\alpha_1\alpha_2\alpha_3}{\alpha_4^3}\right]^{s/24}
\int_0^1 dZ Z^{-S-2+s/3}(1-Z)^{-t-2+s/12}\nonumber\\
&&\left(\frac{\alpha_{12}^2(1-Z)
    +\alpha_{23}^2Z-\alpha_{13}^2Z(1-Z)}{\alpha_4^2}\right)^{-s/12}\nonumber\\
&&\left[\left|\frac{x_1(x_1-Z)(x_2-1)}{x_2(x_2-Z)(x_1-1)}\right|^{1/4}
+\left|\frac{x_2(x_2-Z)(x_1-1)}{x_1(x_1-Z)(x_2-1)}\right|^{1/4}\right]^{s/2}
\label{delta0amp1}
\eea
The second  simple example is
\bea
{\bfs\pi}_1=-{\bfs\pi}_2={\bfs\gamma},\qquad {\bfs\pi}_4=-{\bfs\pi}_3
=-{\bfs\gamma},\qquad {\bfs\gamma}_1=-{\bfs\gamma}_2
\eea
for which the scattering amplitude is
\bea
{\cal M}^{\Delta h=0}_2&=&\left[\frac{\alpha_1\alpha_2\alpha_3}{\alpha_4^3}\right]^{s/24}
\int_0^1 dZ Z^{-S-2+s/12}(1-Z)^{-t-2+s/12}\nonumber\\
&&\left(\frac{\alpha_{12}^2(1-Z)
    +\alpha_{23}^2Z-\alpha_{13}^2Z(1-Z)}{\alpha_4^2}\right)^{-s/12}\nonumber\\
&&\left[\left|\frac{x_1(x_2-Z)(x_1-1)}{x_2(x_1-Z)(x_2-1)}\right|^{1/4}
+\left|\frac{x_2(x_1-Z)(x_2-1)}{x_1(x_2-Z)(x_1-1)}\right|^{1/4}\right]^{s/2}
\label{delta0zmp2}
\eea
Both of these examples assume the individual ${\bfs\pi}$'s have
uniform signs for their individual components. The full range
of ${\bfs\pi}$ conserving
amplitudes for minimal mass external strings is considerably larger.

For this second example, the total compactified momentum in the $S$
channel and in the $t$ channel are both zero. Thus $S\leftrightarrow t$
crossing symmetry reads 
${\cal M}_2(S,t;\alpha_1,\alpha_2,\alpha_3,\alpha_4)={\cal M}_2(t,S;\alpha_3,\alpha_2,\alpha_1,\alpha_4)$. This can be proved by changing integration
variables $Z\to 1-Z$ and using the identities (\ref{crossing}).

\section{High energy four string scattering}
In this section we discuss the Regge high energy behavior of
some of the four string amplitudes. In a Poincar\'e invariant
theory the Regge limit is $-S\to\infty$ at fixed $t$ and
the amplitudes have the typical behavior
\bea
{\cal M}\sim \beta(t)(-S)^{\alpha(t)}.
\eea
The Regge trajectory $\alpha(t)$ passes through nonnegative
integers $J$ when $t$ passes through the squared mass eigenvalues of
the string: $J$ is then the spin of the string mass eigenstate.
Physical scattering requires
$t<0$, whereas the masses squared are nonnegative.
Thus $\alpha(t)$ is an extrapolation of the relation between
angular momentum and string mass squared. In string theory in
tree approximation this
relationship is linear $\alpha(t)=\alpha^\prime t +\alpha_0$.
We have chosen units where $\alpha^\prime=1$.

For $s>0$, the generalized protostring does not enjoy Poincar\'e invariance.
In particular the four string amplitudes
depend on the $\alpha_k=2p^+_k$ in addition to
$S$ and $t$, and one can consider various Regge limits, depending
on whether some of the $\alpha_k$'s are also getting large with $-S$.
We shall be particularly interested in limits where $\alpha_1$ and
$-\alpha_4$ tend to infinity linearly with $-S$, with $\alpha_{2,3}$
fixed. With Poincar\'e invariance, one can always choose a frame where
this is true. This is the frame that Mueller chose to show
that the physics of the lightcone worldsheet for open strings
implies that open string scattering amplitudes
in the forward direction must grow
linearly with $-S$ at fixed $\eta=\alpha_1/(-S)$ \cite{mueller}.
This behavior corresponds to constant cross sections by the
optical theorem.
In a Poincar\'e invariant
theory this implies $\alpha(0)=1$. But in the protostring models
this last conclusion does not necessarily hold, as we shall see.

For the Lorentz covariant bosonic open string 
scattering amplitude ($s=0$), the limit $-S=(p_1+p_2)^2
\to\infty$ is evaluated by changing variables $Z=e^{-u/(-S)}$
\bea
{\cal M}&\approx&(-S)^{t+1}
\int_0^\infty du e^{-u}u^{-t-2}=
(-S)^{t+1}\Gamma(-t-1),\qquad S\to -\infty.
\eea
where $-t=(p_2+p_3)^2$ is the momentum transfer ($=0$ in the forward
direction). Here $\alpha(t)=t+1$ so indeed $\alpha(0)=1$.

When the Grassmann dimension $s>0$, 
the analysis is complicated by the explicit dependence of
the integrand on $x_\pm$. In the following we obtain the high energy
behavior of the scattering amplitudes obtained in Section 3.
\subsection{$\Delta h^a=\pm2$}
Applying the change of variables $Z=e^{-u/(-S)}$ developed in
Section 2.3 to (\ref{delta2amp}) using 
(\ref{xhe}), we find the leading high energy behavior
\bea
{\cal M}^{\Delta h=2}&\approx&(-S)^{t+1-s/12}
\frac{|\alpha_4|^{s/4}}{|\alpha_1\alpha_2\alpha_3|^{s/12}}
\int_0^\infty du e^{-u}u^{-t-2+s/12}\nonumber\\
&&\left|\frac{\alpha_{14}^2+\eta^2u^2+2\eta u(\alpha_2-\alpha_3)}
  {\alpha_4^2}\right|^{s/24}.
\eea
In the limit we are taking $\alpha_4\approx-\alpha_1$ and $\alpha_{14}$,
$\alpha_2$, $\alpha_3$ are fixed. Making these replacements then gives
\bea
{\cal M}^{\Delta h=2}&\approx&(-S)^{t+1-s/12}\alpha_1^{s/12}
\frac{1}{|\alpha_2\alpha_3|^{s/12}}
\int_0^\infty du e^{-u}u^{-t-2+s/12}\nonumber\\
&&\left|{\alpha_{14}^2+\eta^2u^2+2\eta u(\alpha_2-\alpha_3)}\right|^{s/24}
\label{delta2amphe}
\eea
Here we see that if $\alpha_1$ and $-S$ go to infinity at fixed ratio,
the result is in accord with Mueller's argument. Note however, that the
power of $(-S)$, $\alpha(t)=t+1-s/12$, so $\alpha(0)=1-s/12$ is {\it not}
unity, as must be the case if the theory were Poincar\'e covariant.

\subsection{$\Delta h^a=0$}
We next turn to the high energy behavior of the amplitudes
${\cal  M}^{\Delta h=0}_1$
and ${\cal M}^{\Delta h=0}_2$. Making the approximation for high energy
\bea
{\cal M}^{\Delta h=0}_1&\approx&\left[\frac{\alpha_1\alpha_2\alpha_3}{\alpha_4^3}
\right]^{s/24}(-S)^{t+1-s/12}
\int_0^\infty du e^{-u(-S-1+s/3)/(-S)}{u}^{-t-2+s/12}\nonumber\\
&&\left(\frac{\alpha_{23}^2+\eta^2u^2+2(\alpha_2-\alpha_3)\eta u
    }{\alpha_4^2}\right)^{-s/12}\nonumber\\
&&\left[\left|\frac{x_1(x_1-Z)(x_2-1)}{x_2(x_2-Z)(x_1-1)}\right|^{1/4}
+\left|\frac{x_2(x_2-Z)(x_1-1)}{x_1(x_1-Z)(x_2-1)}\right|^{1/4}\right]^{s/2}
\eea
where, identifying $x_1=x_-$ and $x_2=x_+$,
\bea
\left|\frac{x_1(x_1-Z)(x_2-1)}{x_2(x_2-Z)(x_1-1)}\right|&\approx&
\left|\frac{\alpha_{14}-\eta u-
  \sqrt{\alpha_{14}^2+\eta^2u^2+2\eta u(\alpha_2-\alpha_3)
   }}{\alpha_{14}-\eta u+
  \sqrt{\alpha_{14}^2+\eta^2u^2+2\eta u(\alpha_2-\alpha_3)
   }}\right|\\
&&\left|\frac{\alpha_{14}+\eta u+
  \sqrt{\alpha_{14}^2+\eta^2u^2+2\eta u(\alpha_2-\alpha_3)
   }}{\alpha_{14}+\eta u-
  \sqrt{\alpha_{14}^2+\eta^2u^2+2\eta u(\alpha_2-\alpha_3)
  }}\right|\\
&\approx&\left|\frac{\eta u+\alpha_2-\alpha_3+
  \sqrt{\alpha_{14}^2+\eta^2u^2+2\eta u(\alpha_2-\alpha_3)
   }}{\eta u+\alpha_2-\alpha_3-
  \sqrt{\alpha_{14}^2+\eta^2u^2+2\eta u(\alpha_2-\alpha_3)
  }}\right|
\eea
The second example in the high energy limit becomes
\bea
{\cal M}^{\Delta h=0}_2&\approx&\left[\frac{\alpha_1\alpha_2\alpha_3}{\alpha_4^3}
\right]^{s/24}(-S)^{t+1-s/12}
\int_0^\infty du e^{-u(-S-1+s/12)/(-S)}{u}^{-t-2+s/12}\nonumber\\
&&\left(\frac{\alpha_{23}^2+\eta^2u^2+2(\alpha_2-\alpha_3)\eta u
    }{\alpha_4^2}\right)^{-s/12}\nonumber\\
&&\left[\left|\frac{x_2(x_1-Z)(x_2-1)}{x_1(x_2-Z)(x_1-1)}\right|^{1/4}
+\left|\frac{x_1(x_2-Z)(x_1-1)}{x_2(x_1-Z)(x_2-1)}\right|^{1/4}\right]^{s/2}
\eea
where
\bea
\left|\frac{x_2(x_1-Z)(x_2-1)}{x_1(x_2-Z)(x_1-1)}\right|&\approx&
\left|\frac{\eta u+\alpha_2-\alpha_3+
  \sqrt{\alpha_{14}^2+\eta^2u^2+2\eta u(\alpha_2-\alpha_3)
   }}{\eta u+\alpha_2-\alpha_3-
  \sqrt{\alpha_{14}^2+\eta^2u^2+2\eta u(\alpha_2-\alpha_3)
  }}\right|
\eea
because $x_2/x_1\approx x_1/x_2$ in the high energy limit we are discussing.
Indeed the differences between ${\cal M}_2$ and ${\cal M}_1$ in this
high energy
limit are subleading so, in leading order we can write
\bea
{\cal M}^{\Delta h=0}_1&\approx&{\cal M}^{\Delta h=0}_2\nonumber\\
&\approx&\left|\frac{\alpha_2\alpha_3}{\alpha_{23}^2}\right|^{s/24}
\left|\frac{\alpha_{23}}{\alpha_{1}}\right|^{-s/12}(-S)^{t+1-s/12}
\int_0^\infty du e^{-u}{u}^{-t-2+s/12}\nonumber\\
&&\left(\frac{\alpha_{23}^2+\eta^2u^2+2(\alpha_2-\alpha_3)\eta u
    }{\alpha_{23}^2}\right)^{-s/12}\nonumber\\
&&\left[\left|\frac{\eta u+\alpha_2-\alpha_3+
  \sqrt{\alpha_{23}^2+\eta^2u^2+2\eta u(\alpha_2-\alpha_3)
   }}{\eta u+\alpha_2-\alpha_3-
  \sqrt{\alpha_{23}^2+\eta^2u^2+2\eta u(\alpha_2-\alpha_3)
  }}\right|^{1/4}\right.\nonumber\\
 &&\left.+\left|\frac{\eta u+\alpha_2-\alpha_3+
    \sqrt{\alpha_{23}^2
      +\eta^2u^2+2\eta u(\alpha_2-\alpha_3)
   }}{\eta u+\alpha_2-\alpha_3-
  \sqrt{\alpha_{23}^2+\eta^2u^2+2\eta u(\alpha_2-\alpha_3)
  }}\right|^{-1/4}\right]^{s/2}
\eea
where we have rearranged factors and dropped some terms, which are
subleading in the high energy limit we are discussing. Here we see how
the argument for linear growth in $\alpha_1$ is satisfied by
this result:
it is a product 
of the factors $\alpha_1^{s/12}$ and $(-S)^{1+t-s/12}$, which
in the forward direction scales like $\alpha_1$ provided $\eta$ is fixed.

If $|\eta|<<|\alpha_{23}|$ the complicated factors simplify
considerably:
\bea
{\cal M}^{\Delta h=0}_1&\approx&{\cal M}^{\Delta h=0}_2\nonumber\\
&\approx&\left|\frac{\alpha_2\alpha_3}{\alpha_{23}^2}\right|^{s/24}
\left|\frac{\alpha_{23}}{\alpha_{1}}\right|^{-s/12}
\left[\left|\frac{\alpha_2}{\alpha_3}\right|^{1/4}
  +\left|\frac{\alpha_3}{\alpha_2}\right|^{1/4}\right]^{s/2}
(-S)^{t+1-s/12}\Gamma(-t-1+s/12)
\eea
which is in accord with the high energy limit obtained in
\cite{thornprotoamp}, which implicitly assumed very small $\eta$.

Here we see that, with $\alpha_{23}\neq0$ and fixed, the coefficient of 
$\alpha_1^{t+1}$ has poles at $t=n+s/12-1$ which are the mass squared  
eigenvalues of the open generalized protostring.
The linear high energy behavior, when $|\eta|\ll|\alpha_{23}|$,
at $t=0$ is the product of $(-S)^{1-s/12}$ and $\alpha_1^{s/12}$
netting precisely linear growth in the forward direction. 

Contrast this with the high energy limit taken with $\alpha_{23}=0$ from
the beginning:
\bea
{\cal M}^{\Delta h=0}_1&\approx&{\cal M}^{\Delta h=0}_2\nonumber\\
&\approx&\left|{\alpha_2}\right|^{s/12}
(-S)^{t+1}
\int_0^\infty du e^{-u}{u}^{-t-2}
\left({\eta u+4\alpha_2}
  \right)^{-s/12}\nonumber\\
&&\left[\left|\frac{\eta u+2\alpha_2+
  \sqrt{\eta^2u^2+4\eta u\alpha_2
   }}{\eta u+2\alpha_2-
  \sqrt{\eta^2u^2+4\eta u\alpha_2
  }}\right|^{1/4}
+\left|\frac{\eta u+2\alpha_2+
    \sqrt{
      \eta^2u^2+4\eta u\alpha_2
   }}{\eta u+2\alpha_2-
  \sqrt{\eta^2u^2+4\eta u\alpha_2
  }}\right|^{-1/4}\right]^{s/2}
\eea
For $\eta<<\alpha_2$ this result simplifies to
\bea
{\cal M}^{\Delta h=0}_1&\approx&{\cal M}^{\Delta h=0}_2\nonumber\\
&\approx&{4}^{-s/12}(-S)^{t+1}
\int_0^\infty du e^{-u}{u}^{-t-2}={4}^{-s/12}(-S)^{t+1}
\Gamma(-t-1)
\eea
which is just the Regge behavior of the bosonic string amplitude.
We stress that the limit taken here is $\alpha_1, -S\to\infty$
at fixed ratio. The coefficient of the Regge behavior is a function
of $t$. Its pole locations are {\it not} those of the
particles of the theory: they correspond to a linear Regge trajectory of
intercept 1. Because the formula was obtained assuming $\alpha_3=-\alpha_2$,
the high energy behavior comes from the collision of two interaction
points on the lightcone worldsheet, and not from the long time propagation
of a protostring mass eigenstate as in the $\alpha_{23}\neq0$ case.
This mismatch can occur because the Lorentz boost symmetry
generated by $M^{-k}$ is absent in the generalized protostring:
for the protostring because
there is no transverse space and for $0<s<24$ because this part of the
Lorentz symmetry is broken. For the bosonic string ($s=0$), of course, 
there is no such mismatch.

\section{Concluding Remarks}
In \cite{thornprotoamp} $N$ string scattering amplitudes were obtained for
  the generalized protostring. The present article focused on the case $N=4$,
  which was treated only lightly in \cite{thornprotoamp}.
  In this case the conformal mapping used in Mandelstam's interacting string
  formalism can be inverted explicitly, which is not possible for
  higher $N$, In particular the images of the joining/breaking points
  on the lightcone worldsheet are explicit functions of the Koba-Nielsen
  variables. We could then use these explicit formulas to obtain
  explicit integral representations of four string tree amplitudes.

  We interpreted the zero mode of the worldsheet momentum density for
  the compactified coordinates representing the spinor world sheet
  fields as ``helicity'' $h^a$, $a=1,2,\ldots, s/2$. and then studied the
  4-string amplitudes for $\Delta h^a=2$ and $\Delta h^a=0$.
  we calculated the pole locations and the Regge high energy behavior
  of the amplitudes.

  The high energy behavior of these models is compatible with Mueller's
  argument that total cross sections for open string scattering approach
  constants in the limit. In Lorentz covariant theories, this would imply
  a Regge trajectory with intercept $\alpha(0)=1$, and the existence of
  massless spin one particles. In the amplitudes studied in the present
  article the mass spectrum implies $\alpha(0)=1-s/12$, but for
  $s>0$ Lorentz covariance is lost. Mueller's argument chooses a frame
  in which $(-S)\propto p^+_1$ and uses the physics of the lightcone
  worldsheet to show that as $p^+_1\to \infty$ the cross section
  approaches a nonzero constant.  Since the models studied here 
  are not Lorentz covariant, the constant cross sections implied by
  Mueller's argument are realized by a behavior $(p^+_1)^{1-\alpha(0)}
  (-S)^{\alpha(0)}$. If $(-S)$ tends to $\infty$ at fixed $p^+_k$,
  cross sections need not tend to constants.
  
\vskip18pt
\noindent\underline{Acknowledgments}: 
This research was supported in part by the Department
of Physics at the University of Florida.
\appendix
\section{The Scattering Amplitudes}
We quote the formulas for the general $N$-string scattering amplitudes
obtained in \cite{thornprotoamp}. They are expressed as integrals over
Koba-Nielsen variables $Z_k$, $k=1,\ldots,N$ and the mapping function
from the Koba-Nielsen plane (or half plane) $z$ to the lightcone worldsheet
\cite{mandelstamlc}.
\bea
\rho&=&\sum_{k=1}^{N-1}\alpha_k\ln(z-Z_k),
\label{rhoofzo}
\eea
where $\alpha_k=2p^+_k\equiv\sqrt{2}(p^0+p^{d+1})$ is positive (negative) if
the external string is incoming (outgoing).
We also introduce the quantities $x_r$, which are the zeros of
$d\rho/dz$.
\bea
\frac{d\rho}{dz}\bigg|_{z=x_r}=0
\eea
and depend on the $Z$'s and $\alpha$'s.
For convenience we choose $Z_1=0$, $Z_{N-1}=1$, and $Z_N=\infty$. The $x_r$
are roots of a polynomial of order $N-2$ so we can take
$r=1,2,\ldots,N-2$.

The kinematics of the scattering process are determined by the
incoming momenta
of the external strings. Of these there are the momenta of the
$d+2$ uncompactified coordinates $p_k^\mu$, which are continuous and conserved
($\sum_k p_k^\mu=0$). After bosonizing the Grassmann worldsheet fields
there are also the $s/2$ dimensional discretized momenta
${\bfs\pi}_k$ which are not conserved:
Their components are positive or negative
odd multiples of a number $\gamma$ which is determined by
insisting that the three string  vertex, initially defined in terms
of string bits, has a finite continuum limit.

The $x_r$ are the images in the Koba-Nielsen plane of the breaking or joining
points on the lightcone worldsheet. The non conservation of the
${\bfs\pi}_k$ is specified by assigning a spurious discretized momentum
${\bfs\gamma}_r$ to each vertex. Each of the $s/2$ components of
${\bfs\gamma}_r$ are $\pm\gamma$, and the full vertex is obtained by
summing over all choices of signs. For each choice, the non-conservation
of discretized momenta reads.
\bea
\sum_k{\bfs\pi}_k+\sum_r{\bfs\gamma}_r=0
\eea
Note that for even $N$ some of the choices satisfy $\sum_r{\bfs\gamma}_r=0$
for which $\sum_k{\bfs\pi}_k=0$.
Mandelstam's interacting string formalism
\cite{mandelstamlc} works for any value of $\gamma$ and gives
for the Koba-Nielsen integrand of the open string amplitudes
\bea
{\cal I}_O&=&\prod_{k=1}^N{1\over \sqrt{|
\alpha_k|}}\left[{\prod_{k<N}|\alpha_k|\over
|\alpha_N|}\right]^{s(1-8\gamma^2)/32}\left[{\prod_{r<t}|x_t-x_r|
\prod_{m<l}|Z_l-Z_m|
\over \prod_{l,r}|Z_l-x_r|}\right]^{s/48}\nonumber\\
&&(2\delta)^{s(N-2)\gamma^2/4}\left[{\prod_{r<s}|x_r-x_s|^{2{\bfs\gamma}_r\cdot{\bfs\gamma}_s
-s\gamma^2/2}
\prod_{k< l<N}|Z_k-Z_l|^{2P_k\cdot P_l-s\gamma^2/2}
\over\prod_{r,k<N}|x_r-Z_k|^{-2{\bfs \pi}_k{\bfs\gamma}_r-s\gamma^2/2}}\right]
\eea
In string bit models, the factor within the first set of square brackets would scale as $M^{N-2}$
where $M$ is the bit number. So
a finite continuum limit requires $\gamma^2=1/8$, in which case
\bea
{\cal I}_O&=&(2\delta)^{s(N-2)/32}\prod_{k=1}^N{1\over \sqrt{|
\alpha_k|}}\left[{\prod_{r<s}|x_r-x_s|^{2{\bfs\gamma}_r\cdot{\bfs\gamma}_s
-s/24}
\prod_{k< l<N}|Z_k-Z_l|^{2P_k\cdot P_l-s/24}
\over\prod_{r,k<N}|x_r-Z_k|^{-2{\bfs\pi}_k{\bfs\gamma}_r-s/24}}\right]
\label{openintegrand}
\eea
Applying parallel considerations to the closed string leads to
\bea
{\cal I}_C&=&\prod_{k=1}^N{1\over{|
\alpha_k|}}\left[{\prod_{k<N}|\alpha_k|\over
|\alpha_N|}\right]^{s/16-s\gamma^2/8}\left[{\prod_{r<t}|x_t-x_r|
\prod_{m<l}|Z_l-Z_m|
\over \prod_{l,r}|Z_l-x_r|}\right]^{s/24}\nonumber\\
&&(4\delta)^{s(N-2)\gamma^2/8}\left[{\prod_{r<s}|x_r-x_s|^{{\bfs\gamma}_r\cdot{\bfs\gamma}_s-s\gamma^2/4}
\prod_{k< l<N}|Z_k-Z_l|^{P_k\cdot P_l-s\gamma^2/4}
\over\prod_{r,k<N}|x_r-Z_k|^{-{\bfs\pi}_k\cdot{\bfs\gamma}_r-s\gamma^2/4}}
\right]
\eea
and a smooth continuum limit in the closed case requires $\gamma^2=1/2$:
\bea
{\cal I}_C&=&(4\delta)^{s(N-2)/16}\prod_{k=1}^N{1\over{|
\alpha_k|}}\left[{\prod_{r<s}|x_r-x_s|^{{\bfs\gamma}_r\cdot{\bfs\gamma}_s
-s/12}
\prod_{k< l<N}|Z_k-Z_l|^{P_k\cdot P_l-s/12}
\over\prod_{r,k<N}|x_r-Z_k|^{-{\bfs\pi}_k\cdot{\bfs\gamma}_r-s/12}}\right]
\label{closedintegrand}
\eea
In these formulas each component of ${\bfs\gamma_r}$ is
$\pm\gamma=\pm1/(2\sqrt{2})$ for the open string and $\pm\gamma=\pm1/\sqrt{2}$
for the closed string.
And of course each component of ${\bfs \pi}_k$ is an odd integer multiple of
$\gamma$.

The scattering amplitudes are obtained by integrating the expression
(\ref{openintegrand}) or (\ref{closedintegrand})
over the unfixed $Z_k$. In the case of the open string, the $Z_k$ are
on the real axis satisfying $Z_1=0<Z_2<\cdots<Z_{N-2}<Z_{N-1}=1$. In the
case of the closed string the $Z_k$ for $k=2,\ldots N-2$ are integrated
over the whole complex plane. In both cases $Z_1=0, Z_{N-1}=1, Z_N=\infty$.
We remind the reader that for physical values of the momenta the resulting
integrals are formally divergent. To handle these divergences, one
starts with (unphysical) values of the momenta for which the integrals
converge, and then one analytically continues to the physical values.
For open string amplitudes one can do this keeping the range of
the $Z$ integrations complete. But for closed string amplitudes one
is forced to divide the integration region up into cells, with
separate analytic continuations in each cell.
\subsection{Maximal helicity violation}
A dramatic simplification occurs when there
is maximal helicity violation. For instance, choose all components of
the first $N-1$ ${\bfs \pi}_k$ to have the value $-\gamma$.
Then necessarily all components of ${\bfs \pi}_N$ and of
each ${\bfs\gamma_r}$  have the value $+\gamma$.  In this case 
${\bfs\gamma}_r\cdot{\bfs\gamma_s}={\bfs \pi}_k\cdot{\bfs \pi}_l
=-{\bfs\gamma}_r\cdot{\bfs \pi}_l=s\gamma^2/2$ for $k,l\neq N$ and 
${\bfs \pi}_N$ doesn't appear in the formula. Then for
the open case ($\gamma^2=1/8$) the 
contribution to the integrand of the scattering amplitude is
\bea
&&\hskip-1.5cm I_O\left|{\partial T\over\partial Z}\right| 
{\det}^{-(24-s/2)/2}(-\nabla^2)_{\rm open}\nonumber\\
&\to&(2\delta)^{s(N-2)/32}\prod_{k=1}^N{1\over \sqrt{|
\alpha_k|}}\left[{\prod_{r<s}|x_r-x_s|^{2}
\prod_{k< l<N}|Z_k-Z_l|^{2}
\over\prod_{r,k<N}|x_r-Z_k|^{2}}\right]^{s/24}|Z_k-Z_l|^{p_k\cdot p_l}\\
&\to&\epsilon^{s(N-2)/32}\prod_{k=1}^N{1\over \sqrt{|
\alpha_k|}}\frac{|
\alpha_N|^{s(N-1)/12}}{\prod_{k<N}|\alpha_k|^{s/12}}
\left[\frac{\prod_{r<s}|x_r-x_s|^{s/12}}
{\prod_{k< l<N}|Z_k-Z_l|^{s/12}}\right]|Z_k-Z_l|^{p_k\cdot p_l}\eea
where we made use of 
Eq.(128) (Eq(B3) of the PRD version) of \cite{thornprotoamp} 
to arrive at the last line.
With this simple choice the scattering amplitude is then
\bea
A^{\rm Open}_N&=&g^{N-2}\prod_{k=1}^N{1\over \sqrt{|\alpha_k|}}
\frac{|\alpha_N|^{s(N-1)/12}}{\prod_{k<N}|\alpha_k|^{s/12}}
\int dZ_2\cdots dZ_{N-1}\nonumber\\
&&{\prod_{r<s}|x_r-x_s|^{s/12}}
{\prod_{k< l<N}|Z_k-Z_l|^{2p_k\cdot p_l-s/12}}
\label{equalgammao}
\eea
where we identified $g=[2\delta]^{s/32}$.
Making the same simplifications for the case of the closed string leads to
\bea
A^{\rm Closed}_N
&=&G^{(N-2)}\prod_{k=1}^N{1\over{|
\alpha_k|}}\frac{|
\alpha_N|^{s(N-1)/6}}{\prod_{k<N}|\alpha_k|^{s/6}}
\int d^2Z_2\cdots d^2Z_{N-1}\nonumber\\
&&{\prod_{r<s}|x_r-x_s|^{s/6}}
{\prod_{k<l<N}|Z_k-Z_l|^{p_k\cdot p_l-s/6}}
\label{equalgammac}
\eea
where $G=2^{s/16}g^2$.
If desired, one can replace $2p_k\cdot p_l$ by $(p_k+p_l)^2-2+s/6$ 
in the open case
and by $(p_k+p_l)^2-8+2s/3$ in the closed case.
Keep in mind that this simpler expression only applies for a very special
choice for the ${\bfs\pi}$'s and ${\bfs\gamma}$'s.
In particular, even for a particular set of 
the ${\bfs\pi}_k$, the full insertion factor at each vertex 
is $2\cos(\gamma\phi)$, which
can be implemented by summing the amplitudes over each component of
each ${\bfs \gamma}_r$ assuming both possible values $\pm\gamma$.

\end{document}